\documentclass[conference]{IEEEtran}
\IEEEoverridecommandlockouts

\usepackage{cite}
\usepackage{amsmath,amssymb,amsfonts}
\usepackage{algorithmic}
\usepackage{graphicx}
\usepackage{textcomp}
\usepackage{xcolor}
\usepackage{url}
\usepackage{doi} 
\usepackage{booktabs}
\usepackage{makecell}
\usepackage[a4paper, left=15mm, right=15mm, top=20mm, bottom=45mm]{geometry}
\usepackage{xcolor}
\usepackage{multirow}
\usepackage{array}       
\usepackage{makecell}    
\usepackage{stfloats}
\def\BibTeX{{\rm B\kern-.05em{\sc i\kern-.025em b}\kern-.08em
    T\kern-.1667em\lower.7ex\hbox{E}\kern-.125emX}}
\begin{document}

\title{Software Defined Vehicle Code Generation: \\
A Few-Shot Prompting Approach
}

\author{\IEEEauthorblockN{1\textsuperscript{st} Quang-Dung Nguyen}
\IEEEauthorblockA{\textit{University of Information Technology} \\
\textit{VNUHCM}\\
Ho Chi Minh City, Vietnam \\
22520286@gm.uit.edu.vn}
\and
\IEEEauthorblockN{2\textsuperscript{nd} Tri-Dung Tran}
\IEEEauthorblockA{\textit{University of Information Technology} \\
\textit{VNUHCM}\\
Ho Chi Minh City, Vietnam \\
22520293@gm.uit.edu.vn}
\and
\IEEEauthorblockN{3\textsuperscript{rd} Thanh-Hieu Chu}
\IEEEauthorblockA{\textit{University of Information Technology} \\
\textit{VNUHCM}\\
Ho Chi Minh City, Vietnam \\
22520429@gm.uit.edu.vn}
\and
\IEEEauthorblockN{4\textsuperscript{th} Hoang-Loc Tran}
\IEEEauthorblockA{\textit{University of Information Technology} \\
\textit{VNUHCM}\\
Ho Chi Minh City, Vietnam \\
locth@uit.edu.vn}
\and
\IEEEauthorblockN{5\textsuperscript{th} Xiangwei Cheng}
\IEEEauthorblockA{\textit{Ferdinand-Steinbeis-Institut} \\
Heilbronn, Germany \\
chris.cheng@ferdinand-steinbeis-institut.de}
\and
\IEEEauthorblockN{6\textsuperscript{th} Dirk Slama}
\IEEEauthorblockA{\textit{Ferdinand-Steinbeis-Institut} \\
Heilbronn, Germany \\
dirk.slama@ferdinand-steinbeis-institut.de}
}

\maketitle

\begin{abstract}
The emergence of Software-Defined Vehicles (SDVs) marks a paradigm shift in the automotive industry, where software now plays a pivotal role in defining vehicle functionality, enabling rapid innovation of modern vehicles. Developing SDV-specific applications demands advanced tools to streamline code generation and improve development efficiency. In recent years, general-purpose large language models (LLMs) have demonstrated transformative potential across domains. Still, restricted access to proprietary model architectures hinders their adaption to specific tasks like SDV code generation. In this study, we propose using prompts, a common and basic strategy to interact with LLMs and redirect their responses. Using only system prompts with an appropriate and efficient prompt structure designed using advanced prompt engineering techniques, LLMs can be crafted without requiring a training session or access to their base design. This research investigates the extensive experiments on different models by applying various prompting techniques, including bare models, using a benchmark specifically created to evaluate LLMs’ performance in generating SDV code. The results reveal that the model with a few-shot prompting strategy outperforms the others in adjusting the LLM answers to match the expected outcomes based on quantitative metrics.
\end{abstract}


\section{Introduction}

The rapid advancement of Artificial Intelligence (AI) has significantly transformed various industries, including automotive, where vehicles are evolving into intelligent mobility platforms \cite{kamran2022artificial,barron2021road}. Software now plays a central role in defining vehicle functionality and user experience. 

AI-driven design and prototyping is becoming a key trend \cite{bilgram2023accelerating}. Large Language Models (LLMs), trained on massive code and natural language corpora, have shown strong potential for Software-Defined Vehicles (SDVs), enabling rapid feature iteration, reducing human error, and improving code quality \cite{belzner2023large}. However, general-purpose LLMs provide only partial knowledge of automotive-specific APIs and often fail to meet domain requirements. This highlights the role of \textit{Prompt Engineering}, which designs and optimizes prompts to guide LLM outputs more effectively \cite{wang2022no}. Beyond improving accuracy, prompt engineering allows developers to adapt generated code to functional and performance constraints of SDV applications.

Motivated by this gap, this study investigates how prompt engineering can enhance SDV code generation. The contributions of this paper are:  
\begin{itemize}
    \item A prompt engineering solution incorporating multiple techniques for generating executable SDV code through an open-source vehicle API.  
    \item A benchmark specifically constructed to assess LLM performance in SDV code generation.  
    \item A quantitative evaluation of state-of-the-art LLMs (ChatGPT o4-mini-high, Gemini 2.5 Pro) under different prompting modes using representative metrics.  
\end{itemize}

\section{Related Work}

\begin{figure*}[!b]
  \centering  \includegraphics[width=0.95\linewidth]{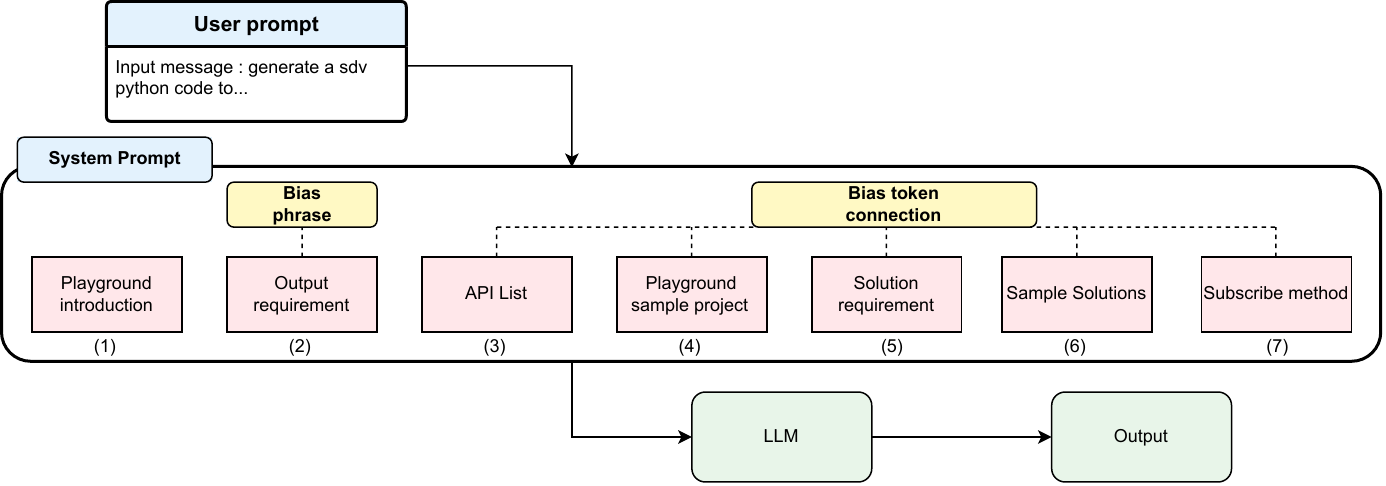}
  \caption{An Overview of our proposed system prompt structure.}
  \label{fig: System prompt design and other LLM components}
\end{figure*}

\subsection{LLM for Code Generation}

LLMs have shown significant potential in automating code generation tasks, offering various levels of specialization and capabilities. These models can be classified based on their architecture, training data, and specific use cases, reflecting the diverse approaches to enhance code generation.

\begin{itemize}
    \item \textbf{General-Purpose LLMs with Code Specialization} are broadly trained models, including on code, enabling tasks like code generation without fine-tuning. Examples like GPT-4 \cite{openai_gpt4} and Bard \cite{google_bard} excel in generating code across languages due to their diverse training datasets.

    \item \textbf{Code-Specific LLMs} are fine-tuned on code-related datasets to enhance tasks like code generation and debugging. Examples include OpenAI's Codex \cite{openai_codex}, derived from GPT-3 and CodeBERT \cite{feng2020codebert}, a bimodal model combining source code and natural language for tasks like code search and generation.

    \item \textbf{Domain-Specific LLMs for Code} are tailored to specific languages, frameworks, or environments. GitHub's CoPilot \cite{github_copilot}, powered by Codex, acts as an AI pair programmer, suggesting code in languages like Python and JavaScript. PyTorch Lightning's Code Generator \cite{pytorch_lightning_code_generator} specializes in generating scripts for machine learning and deep learning tasks in the PyTorch ecosystem.
\end{itemize}

LLMs for code generation automate routine tasks, enhance developer productivity, and support rapid prototyping by generating code from high-level descriptions. Their versatility across multiple languages and frameworks makes them valuable tools for various domains, democratizing software development and enabling broader participation by users with limited programming expertise. Despite these benefits, LLMs have limitations. They may produce syntactically correct but semantically flawed, insecure, or outdated code, requiring expert review. Their high computational costs and inability to debug or refine outputs necessitate significant human intervention.

Prompt engineering addresses these challenges by allowing users to design and refine prompts that guide LLMs effectively. This approach enhances task-specific performance, offering greater control over outputs without modifying the underlying models.

\subsection{Prompt Engineer Techniques}

Brown et al. \cite{Brown_FewShotPrompting} introduced two key prompting techniques: zero-shot and few-shot prompting. Zero-shot prompting, the simplest method, generates responses based solely on a natural language prompt without any provided examples. This approach leverages the model's pre-trained knowledge and ability to generalize from the prompt. While effective for straightforward tasks, it often struggles with more complex or nuanced tasks due to the lack of explicit guidance.

To address these limitations, Few-shot prompting incorporates a small number of task-specific examples within the prompt. This technique, validated in studies on models like GPT-3, enhances the model's ability to generate accurate outputs by providing contextual guidance. By bridging the gap between zero-shot and fully supervised learning, few-shot prompting improves the model's performance on various tasks, including code generation, by making it more aware of the desired output format and structure.

Another advanced prompting technique is the Chain-of-Thought (CoT) prompting \cite{Wei_CoT}, which encourages the model to generate intermediate reasoning steps before arriving at the final answer. This method has been shown to enhance the reasoning abilities of LLMs, such as GPT-3, LaMDA \cite{thoppilan2022lamda}, PaLM \cite{anil2023palm}, and others, by breaking down complex tasks into smaller, more manageable components. By eliciting a sequence of thought processes, CoT prompting improves the model's performance on tasks that require logical reasoning, structured thinking, and detailed step-by-step explanations, making it highly effective for intricate code generation tasks.

AceCoder \cite{AceCoder} offers a specialized approach to code generation using advanced LLMs like GPT-3.5 \cite{openai2023gpt35}, CodeGeex-13B \cite{zheng2023codegeex}, and CodeGen-6B \cite{nijkamp2022codegen}. It employs guided prompts tailored to coding tasks and datasets like MBPP \cite{austin2021program} and MBJP \cite{athiwaratkun2023multilingual}. By utilizing metrics such as Pass@k, correctness, and maintainability, AceCoder ensures high-quality, efficient, and maintainable code generation, showcasing the power of prompt engineering in AI-driven development.

In recent years, SDVs have been reshaping the automotive industry, with vehicle functionality increasingly reliant on software-controlling hardware components via APIs. This shift demands generating complex, API-integrated code that meets stringent safety, performance, and reliability standards. LLMs hold significant potential for simplifying this process. By applying prompt engineering, developers can guide LLMs to produce precise, efficient code, reducing the time and expertise required. This approach addresses the challenges of SDV development, offering scalable and robust solutions for next-generation vehicles.

\section{Proposed Method}

\begin{figure*}[!b]
  \centering  \includegraphics[width=1\linewidth]{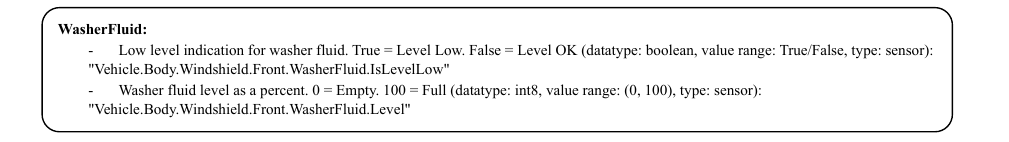}
  \caption{API Description Sample.}
  \label{api description}
\end{figure*}

This research exemplifies dual-purpose scientific inquiry in computer science, tackling real-world problems while producing the research outcomes according to specific case \cite{hevner2010design}. To develop our solution, we followed the CRISP-DM(Q) according to \cite{studer2021towards} as follows: A detailed understanding of the business and data is given by close exchange with Automotive experts in working groups, In terms of Data Preparation, we focused on retrieving and selecting the vehicle APIs related to automotive application development. Since the method does not originally fine-tune a pretrained-model, the phase Modeling is based on applying prompt engineering with multiple techniques (mainly few-shot) to guide LLMs in generating SDV code. To determine the feasibility and superiority of the solution, the proposed solution was deployed on an online prototyping platform and compared with other existing solutions and the original LLMs based on our SDV benchmark (Evaluation phase). The SDV Copilot is still deployed online and steadily monitored, so the stages of Deployment, Monitoring \& Maintenance are also covered.


While fine-tuning is a common way to adapt LLMs, it is limited by proprietary restrictions and hardware demands. Prompting offers a more practical alternative \cite{10.24963/ijcai.2023/577}, enabling precise outputs with suitable design.
Here, the system prompt implementation includes VSS APIs, platform setup, and prompt engineering strategies for SDV prototyping is defined.

\subsection{Vehicle Signal Specification and digital.auto Playground}

VSS is a standard introduced by the Connected Vehicle Systems Alliance (COVESA) to provide a unified, structured representation of vehicle data signals in SDVs. VSS defines both a syntax and a catalog of vehicle signals—covering attributes, actuators, and sensors—commonly found across modern vehicles.

The VSS API enables programmatic access to these vehicle signals through standardized methods, facilitating consistent interaction patterns across different vehicle systems.

The structure of VSS is grounded in a semantic and logical taxonomy, consisting of the following components:

\begin{itemize}
    \item \textbf{Rule Set:} Defines constraints and semantics for signal usage.
    \item \textbf{Data Definitions:} Describes types and hierarchies of vehicle signals.
    \item \textbf{Serialization Tools:} Provides encoding schemes (e.g., JSON, Protobuf) for communication and integration.
\end{itemize}

The digital.auto initiative provides an in-browser Playground that supports rapid prototyping of SDV features using VSS APIs, Python, and JavaScript. Its ProtoPilot tool enables prompt-based interaction with generative AI models from major cloud providers.

The proposed solution aims to leverage the VSS API and Playground to establish a realistic interface for SDV code generation, allowing LLMs to bridge natural language prompts with vehicle signal control.

\subsection{System prompt}

Prompts can be user or system. User prompts interactively adjust model behavior, while system prompts guide it consistently throughout the process \cite{wang2023review}.

In our approach, system prompts embed VSS APIs and few-shot examples to ensure accurate API usage and Python-compliant code generation, yielding cohesive and functional SDV projects.

The work and design of the system prompt and other components of the SDV generator LLM can be visualized as shown in \textbf{Fig. \ref{fig: System prompt design and other LLM components}}. Foundation LLMs exhibit limited knowledge of VSS technical coding, necessitating a system prompt with distinct sections to address specific aspects. The prompt is organized into three key components:

\begin{itemize}
    \item \textbf{Descriptive prompts:} Provide detailed information about digital.auto Playground, proper implementation of VSS APIs in SDV projects, and the expected paradigm of the final output, all presented in natural language.
    \item \textbf{API listing prompts:} Contain information on all supported APIs, serving as an extended knowledge base directly incorporated into the system prompt.
    \item \textbf{I/O examples prompts:} Include response demonstrations that leverage the advantages of few-shot and CoT techniques to enhance the understanding of SDV code and reduce hallucinations.
\end{itemize}

\begin{table*}[!t]
  \small
  \caption[Metric Evaluation Results]{%
    Metric Evaluation Results for LLMs applying few-shot, zero-shot, and baseline models (without system prompt)
  }
  \setlength{\tabcolsep}{5pt}
  \centering
  \renewcommand{\arraystretch}{0.9}
  \begin{tabular}{llcccccccc}
    \toprule
    \textbf{Model} & \textbf{Technique} 
    & \multicolumn{2}{c}{\textbf{CodeBLEU}} 
    & \multicolumn{2}{c}{\textbf{CodeBERTScore}} 
    & \multicolumn{2}{c}{\textbf{ROUGE-L}} 
    & \multicolumn{2}{c}{\textbf{ChrF}} \\
    \cmidrule(lr){3-4} \cmidrule(lr){5-6} \cmidrule(lr){7-8} \cmidrule(lr){9-10}
    & & Score & p-value & $F_{lcs}$ ($\beta=1$) & p-value 
    & $\beta=1$ & p-value & $\beta=1$ & p-value \\
    \toprule
    \multirow{3}{*}{\textbf{Gemini 2.5 Pro}} & Few-shot 
    & \textbf{\textcolor[RGB]{0,115,0}{0.491}} & 1.37e-38 & \textbf{\textcolor[RGB]{0,115,0}{0.717}} & 3.65e-32 & \textbf{\textcolor[RGB]{0,115,0}{0.501}} & 8.41e-25 & \textbf{\textcolor[RGB]{0,115,0}{0.584}} & 4.24e-24 \\
    & Zero-shot 
    & 0.34 & 2.17e-45 & 0.596 & 8.98e-45 & 0.309 & 1.24e-40 & 0.404 & 3.12e-39 \\
    & Original 
    & 0.284 & 5.63e-45 & 0.496 & 5.13e-42 & 0.16 & 1.67e-59 & 0.232 & 2.07e-57 \\
    \midrule
    \multirow{3}{*}{\textbf{ChatGPT o4-mini-high}} & Few-shot 
    & \textbf{\textcolor[RGB]{0,115,0}{0.487}} & 6.06e-32 & \textbf{\textcolor[RGB]{0,115,0}{0.725}} & 4.01e-27 & \textbf{\textcolor[RGB]{0,115,0}{0.553}} & 1.95e-22 & \textbf{\textcolor[RGB]{0,115,0}{0.627}} & 2.86e-21 \\
    & Zero-shot 
    & 0.299 & 1.08e-38 & 0.619 & 1.06e-31 & 0.405 & 1.34e-25 & 0.479 & 1.42e-27 \\
    & Original 
    & 0.253 & 9.70e-48 & 0.567 & 1.30e-34 & 0.346 & 3.19e-36 & 0.424 & 4.02e-39 \\
    \midrule
    \textbf{Claude 3.7 Sonnet} & Original 
    & 0.291 & 7.16e-47 & 0.522 & 1.92e-37 & 0.22 & 8.48e-39 & 0.29 & 3.32e-38 \\
    \bottomrule
  \end{tabular}
  \label{tab:fscore}
\end{table*}

\subsection{Prompt Engineering Application}

Without explicit instructions, LLMs such as ChatGPT often perform poorly in complex tasks. System prompts with few-shot examples (environment details, APIs, use cases) improve accuracy, while adding CoT guidance refines reasoning. For SDV development, we propose a combined strategy that leverages these techniques for precise, context-aware code generation.

In our setup, we distinguish three prompting modes: 
(i) \textbf{Few-shot}, where the full file context is provided together with representative input–output examples; 
(ii) \textbf{Zero-shot}, where the file context remains but code examples are removed; and 
(iii) \textbf{Original}, where no file context is supplied and the model relies solely on its pre-trained knowledge. 
This classification establishes a clear basis for evaluating the role of contextual information in SDV code generation.

The system prompt builds on a zero-shot foundation, detailing key aspects of the SDV project and Playground (e.g., vehicle optimization, context-aware functions, coding syntax) and integrating 745 APIs for broad coverage (\textbf{Fig. \ref{api description}}). On top of this, few-shot prompting adds curated examples for syntactic accuracy and contextual alignment \cite{ijcai2024p702}, while Tree-of-Thought (ToT) explores diverse solutions and reduces bias \cite{NEURIPS2023_271db992}. In total, 22 few-shot and ToT problems are embedded. Chain-of-Thought (CoT) is further integrated via annotated examples, enhancing reasoning for complex tasks.

\subsection{Bias}

Bias is inherent in prompt design, especially in large, context-rich system prompts. LLMs often struggle with coherence due to varied token intent \cite{sieker-etal-2023-beyond}, while vague or poorly chosen phrases may reduce accuracy. Yet bias can also help maintain coherence, particularly in natural language prompts.

In our design, long context texts are segmented, with bias reinforcing semantic connections. For very large prompts, where some instructions lack clarity, iteration bias can improve interpretation \cite{ijcai2024p720}. Here, bias is reflected through key tokens (e.g., API, method) and meaningful phrases (e.g., output bind).

\subsection{Dataset}

A benchmark dataset was constructed to evaluate LLMs in SDV Python code generation by curating 17 representative use cases from the digital.auto project. Each was implemented at three complexity levels—simple, moderate, and advanced—yielding 51 prompt–solution pairs with ground truth authored by senior automotive developers for correctness and realism. This multi-level design ensures a balanced, reproducible benchmark for systematically assessing LLM performance under increasing functional and systemic complexity.

The benchmark dataset introduced in this section has been released publicly at our GitHub repository. \footnote{\url{https://github.com/DungQuangUiT/SDV-Gode-Generation-Benchmark/releases/tag/v1.0}}

\subsection{Metrics}

A combination of syntactic, semantic, and functional metrics, which provide comprehensive insight into the quality of generated code, was employed to evaluate the performance of SDV code generation models. The used metrics are as follows:
\begin{itemize}
    \item \textbf{CodeBLEU}: An extension of BLEU that integrates n-gram matching, abstract syntax tree similarity, and dataflow semantics for comprehensive code quality assessment \cite{ren2020codebleu}.
    \item \textbf{CodeBERTScore}: A metric leveraging CodeBERT embeddings to measure semantic similarity between generated and reference code \cite{zhou-etal-2023-codebertscore}.
    \item \textbf{ROUGE-L}: Evaluates overlap via the Longest Common Subsequence (LCS), reporting precision, recall, and F-score \cite{lin-2004-rouge}.
    \item \textbf{ChrF}: A character n-gram F-score balancing precision and recall, with $\beta=1$ assigning equal weight \cite{popovic-2015-chrf}.

\end{itemize}

\section{Results and analysis}

Evaluation was conducted on ChatGPT o4-mini-high, Gemini 2.5 Pro, and Claude 3.7 Sonnet under different prompt configurations, using the benchmark dataset. Since no domain-specific models exist for SDV code generation, baseline LLMs often failed to produce executable outputs. Minor adjustments were applied to regenerate valid code before evaluating results with the selected metrics.

\textbf{Table \ref{tab:fscore}} shows model performance with equal weighting of recall and precision. The few-shot prompt markedly improved SDV code generation for both ChatGPT o4-mini-high and Gemini 2.5 Pro. Without system prompts, outputs were largely unusable, while zero-shot prompting offered broader but weaker results compared to few-shot. The consistently low p-values confirm that these improvements are statistically significant.

\begin{table}
  \small
  \caption[CodeBLEU Submetric Scores]{%
    CodeBLEU Submetric Scores: Syntax and Dataflow
  }
  \setlength{\tabcolsep}{11pt} 
  \centering
  \renewcommand{\arraystretch}{0.9}
  \begin{tabular}{>{\raggedright\arraybackslash}p{2cm}lcc}
    \toprule
    \textbf{Model} & \textbf{Technique} & \textbf{Syntax} & \textbf{Dataflow} \\
    \toprule
    \multirow{3}{*}{\makecell[l]{\textbf{Gemini}\\\textbf{2.5 Pro}}} & Few-shot 
    & \textbf{\textcolor[RGB]{0,115,0}{0.79}} & \textbf{\textcolor[RGB]{0,115,0}{0.53}} \\
    & Zero-shot 
    & 0.64 & 0.458 \\
    & Original 
    & 0.513 & 0.435 \\
    \midrule
    \multirow{3}{*}{\makecell[l]{\textbf{ChatGPT}\\\textbf{o4-mini-high}}} & Few-shot 
    & \textbf{\textcolor[RGB]{0,115,0}{0.754}} & \textbf{\textcolor[RGB]{0,115,0}{0.555}} \\
    & Zero-shot 
    & 0.579 & 0.426 \\
    & Original 
    & 0.466 & 0.356 \\
    \midrule
    \textbf{Claude 3.7 Sonnet} & Original 
    & 0.521 & 0.452 \\
    \bottomrule
  \end{tabular}
  \label{tab:codeBleu_component}
\end{table}

\begin{table}
  \small
  \caption[Evaluation on LLM applying different prompt components]{%
    Evaluation with different prompt components on Gemini 2.5 Pro
  }
  \setlength{\tabcolsep}{1pt}
  \centering
  \renewcommand{\arraystretch}{0.6}
  \begin{tabular}[t]{@{}lcccc@{}}
    \toprule
      & \textbf{CodeBLEU} & \textbf{CodeBERTScore} & \textbf{ROUGE-L} & \textbf{ChrF} \\
      & CodeBLEU & F1 Score & $F_{lcs}$ ($\beta=1$) & $\beta=1$ \\
    \toprule
    \makecell[l]{\textbf{Full prompt}}     & \textbf{\textcolor[RGB]{0,115,0}{0.491}} & \textbf{\textcolor[RGB]{0,115,0}{0.717}} & \textbf{\textcolor[RGB]{0,115,0}{0.501}} & \textbf{\textcolor[RGB]{0,115,0}{0.584}} \\
    \makecell[l]{\textbf{Without prompt}}                          & 0.284 & 0.496 & 0.160 & 0.232 \\
    \makecell[l]{\textbf{Intro (1)}}     & 0.324 & 0.560 & 0.248 & 0.335 \\
    \makecell[l]{\textbf{Req (25)}}     & 0.454 & 0.674 & 0.446 & 0.537 \\
    \makecell[l]{\textbf{API (3)}}       & 0.264 & 0.533 & 0.208 & 0.294 \\
    \makecell[l]{\textbf{Example (467)}}     & 0.429 & 0.626 & 0.343 & 0.417 \\
    \bottomrule
  \end{tabular}
  \label{tab:section_evaluation}
\end{table}

As illustrated in \textbf{Table \ref{tab:codeBleu_component}}, the average syntax and dataflow scores—submetrics of CodeBLEU—reveal the limited understanding of raw LLMs regarding the specific syntax and structure of proper SDV code. The system prompt helps bridge this gap by aligning the model with correct code syntax and functional structure.

\textbf{Table \ref{tab:section_evaluation}} shows the effect of individual prompt sections on Gemini 2.5 Pro. Each part improved performance over the base model, with Section (25) (output requirements + SDV sample) having the largest impact, followed by Section (467) with 22 executable samples. Overall, the few-shot prompt remains the best-performing approach, supported by these sub-prompts.

\begin{figure*}[h]
  \centering
  \includegraphics[width=1\linewidth]{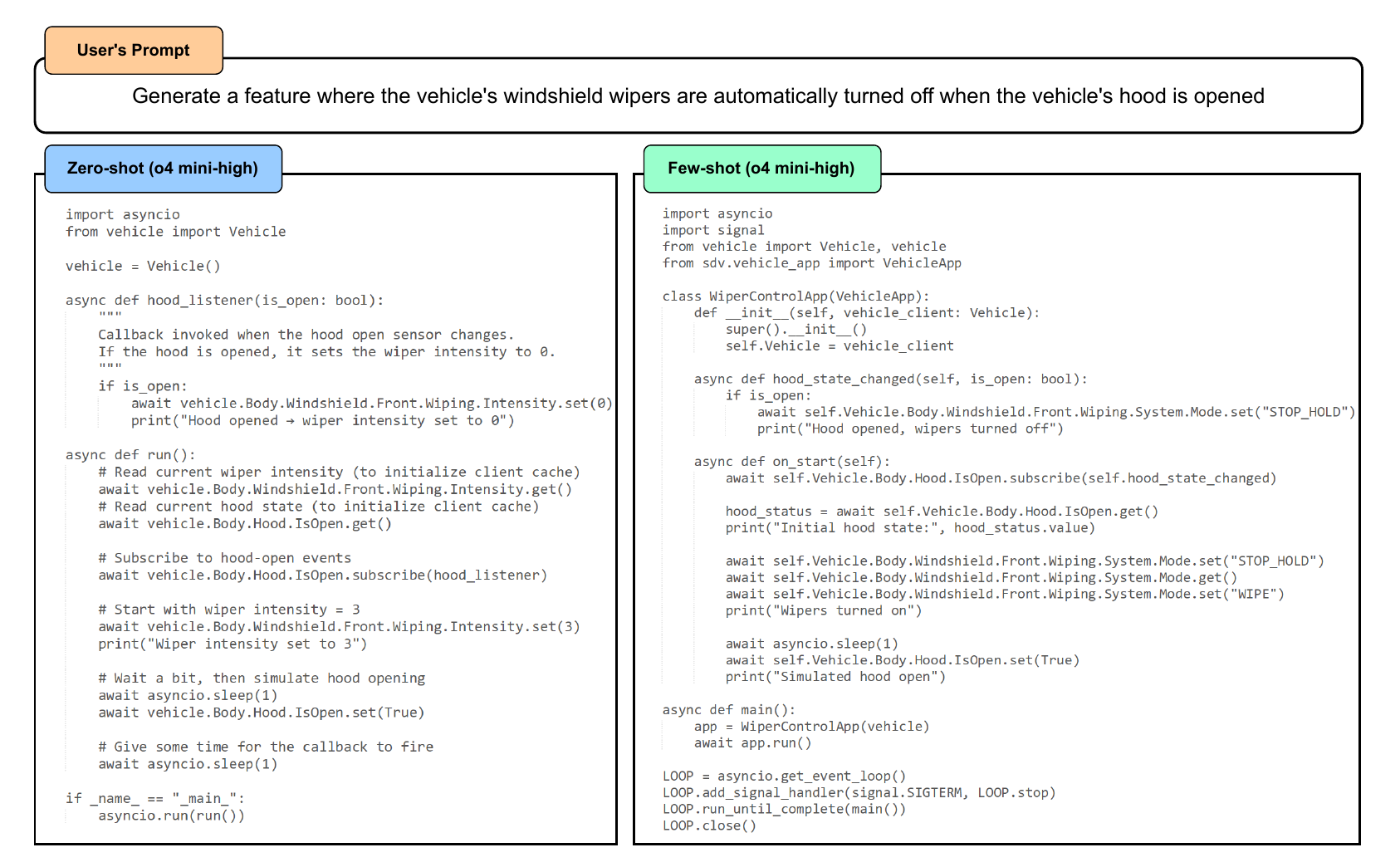}
  \caption{Few-Shot Prompting Delivers Superior Responses Compared to Zero-Shot Prompting}
  \label{fig: zero-shot and few-shot}
\end{figure*}

\textbf{Fig. \ref{fig: zero-shot and few-shot}} compares ChatGPT o4-mini-high outputs. The few-shot version leverages the SDV framework, lifecycle management, and higher-level APIs, yielding executable, domain-aligned code. By contrast, the zero-shot version is procedural, poorly structured, and fails due to incomplete API usage. This highlights the superiority of few-shot prompting for functional, context-aware outputs.

\section{Conclusion}

The prompt engineering solution, combined with multiple techniques, has shown fantastic results to enable SDV application code generation through the COVESA VSS APIs. Experiments show that our system prompt significantly improves LLM performance across quantitative metrics. In addition, we created a benchmark dataset tailored for SDV code generation, offering a valuable resource to support future research in this domain.

While results are promising, several areas remain for improvement. The generalizability of few-shot prompting across diverse APIs and complex scenarios must be further validated. Updating and modifying large system prompts also remains time-consuming. Future work should include expert evaluations to assess the practical usability and quality of generated code, complementing quantitative metrics. Moreover, extending the platform with resources for fine-tuning would enable a more rigorous comparison between prompting strategies and model adaptation methods.

\end{document}